\documentclass[reprint, amsmath, amssymb, aps, superscriptaddress, prb, longbibliography]{revtex4-1}

\usepackage{graphicx}
\usepackage{epstopdf}
\usepackage{dcolumn}
\usepackage{bm}
\usepackage{color}
\usepackage[colorlinks=true, urlcolor=blue, linkcolor=blue, citecolor=blue]{hyperref}
\usepackage{natbib}
\usepackage{amsbsy}
\usepackage{lipsum}
\usepackage{verbatim} 

\usepackage[dvipsnames]{xcolor}
\usepackage{pst-all}
\usepackage{ulem}

\begin{document}

	\title{
		Demonstration of high sensitivity of microwave-induced resistance oscillations to circular polarization
	}	
	
	\author{M.\,L.\,Savchenko}
	\affiliation{Institute of Solid State Physics, Vienna University of
		Technology, 1040 Vienna, Austria}
	\affiliation{Rzhanov Institute of Semiconductor Physics, 630090 Novosibirsk,
		Russia}
	
	\author{A.\,Shuvaev}
	\affiliation{Institute of Solid State Physics, Vienna University of
		Technology, 1040 Vienna, Austria}
	
	\author{I.\,A.\,Dmitriev}
	\affiliation{Terahertz Center, University of Regensburg, 93040 Regensburg,
		Germany}
	
	\author{S.\,D.\,Ganichev}
	\affiliation{Terahertz Center, University of Regensburg, 93040 Regensburg,
		Germany}
	
	\author{Z.\,D.\,Kvon}
	\affiliation{Rzhanov Institute of Semiconductor Physics, 630090 Novosibirsk, 
		Russia}
	\affiliation{Novosibirsk State University, 630090 Novosibirsk, Russia}
	
	\author{A.\,Pimenov}
	\affiliation{Institute of Solid State Physics, Vienna University of
		Technology, 1040 Vienna, Austria}
	
	\begin{abstract}
		\noindent
		We demonstrate that long-debated immunity of microwave-induced
		resistance oscillations (MIRO) to the sense of circular polarization
		is not a generic property of this phenomenon in solid-state
		two-dimensional electron systems. Using a large-area GaAs-based
		heterostructure we detect up to 30 times larger MIRO signal for the
		cyclotron resonance (CR) active helicity, fully consistent with the
		concurrently measured transmission and the deduced CR shape of the
		Drude absorption. We further elaborate conditions to avoid extrinsic
		factors capable of producing an apparent immunity of the
		photoresponse.
	\end{abstract}
	
	\date{\today}
	\maketitle

	In the last two decades, studies of high mobility 2D electron systems (2DES) in a weakly quantizing magnetic field gave access to a family of fascinating interrelated magnetooscillation phenomena reflecting various spatial and spectral resonances that emerge in high Landau levels under application of static and/or alternating (microwave or terahertz) electric fields~\cite{dmitriev:2012,zudov:2001a,mani:2002,zudov:2003,yang:2003,zudov:2001b,yang:2002,zhang:2008, zhang:2007c,wiedmann:2010b,smet:2005,dorozhkin:2011,Bykov2010a,levin:2015,dorozhkin:2016,herrmann:2016,shi:2017}. 
	Experimental research in a growing number of high-mobility 2DES~\cite{konstantinov:2013,zudov:2014,karcher:2016,yamashiro:2015,zadorozhko:2018,otteneder:2018,friess:2020,tabrea:2020,moench:2020,kumaravadivel:2019,savchenko:2020} and various conditions have been accompanied by theoretical developments which provided new insights into the interplay of Landau quantization, disorder, and interactions in electron kinetics in both weakly and strongly nonequilibrium regimes of magnetotransport~\cite{dmitriev:2012,ryzhii:1970,durst:2003,dmitriev:2003,andreev:2003,vavilov:2004,dmitriev:2005,vavilov:2007,dmitriev:2009b,raichev:2008,monarkha:2019,dmitriev:2019,greenaway:2019,raichev:2020,dmitriev:2004,chepelianskii:2009,mikhailov:2011,beltukov:2016}.
	These studies have been largely motivated by the discovery of giant microwave-induced resistance oscillations (MIRO)~\cite{zudov:2001a},  magnetooscillations of photoresistance controlled by the ratio of the microwave and cyclotron frequencies, as well as the radiation-induced zero-resistance states (ZRS)~\cite{mani:2002,zudov:2003,yang:2003} that emerge at deep minima of MIRO and represent a rare example of electric domain instability associated with negative absolute conductivity in strongly driven 2DES.
	
	\begin{figure*}
		\includegraphics[width=2\columnwidth]{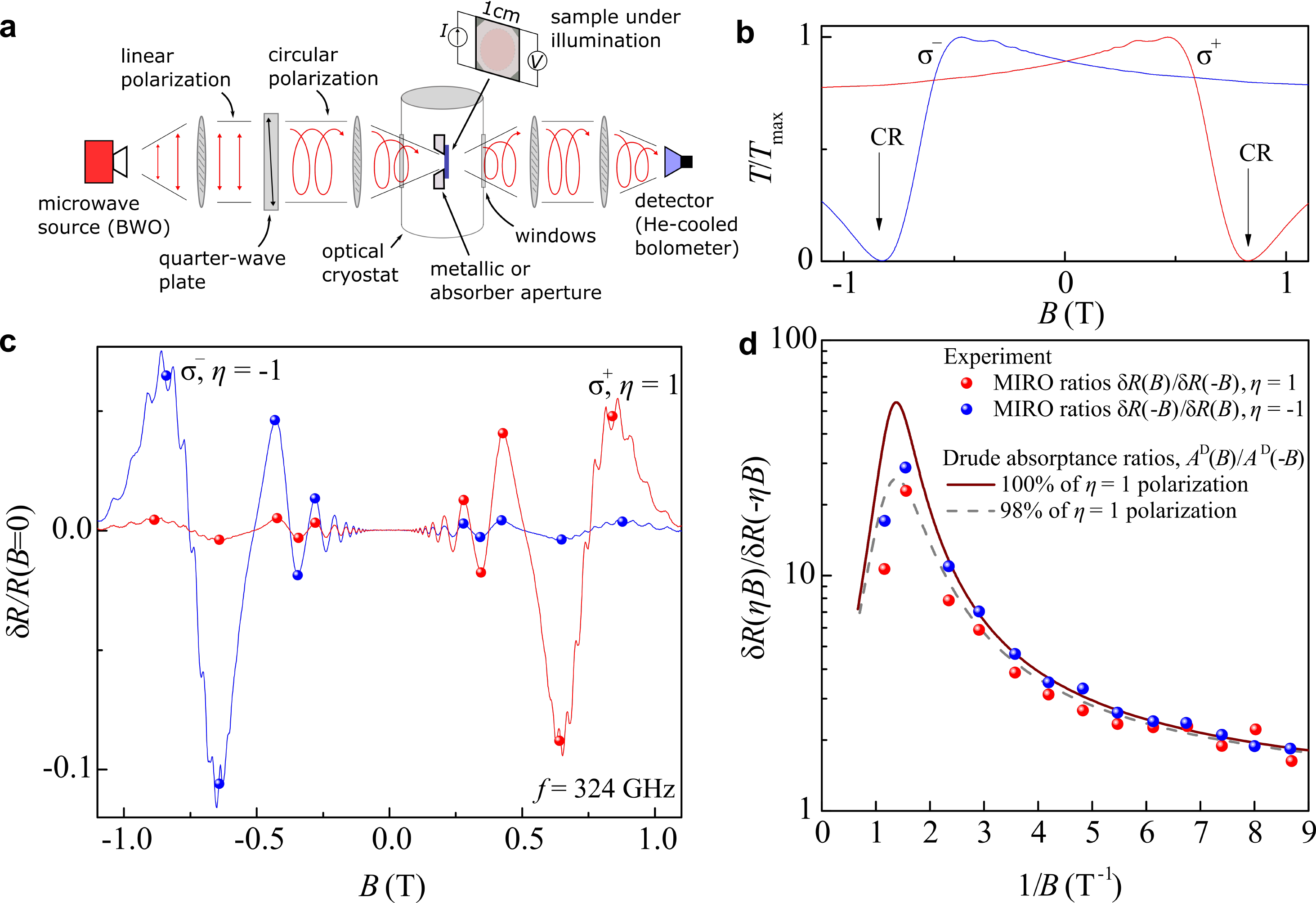}
		\caption{
			\textbf{a},~Scheme of the experiment (the irradiated sample area is highlighted). 
			\textbf{b, c},~Magnetic field dependences of (b) the transmittance $T(B)$, normalized to its maximum value $T_\text{max}$, and (c) MIRO in the longitudinal photoresistance $\delta R$, normalized to the zero field dark resistance $R(B=0)$, for the right-hand ($\sigma^+$, $\eta=1$, red) and left-hand ($\sigma^-$, $\eta=-1$, blue) circularly polarized radiation.
			\textbf{d},~The ratios $\delta R(\eta B)/\delta R(-\eta B)$ at extrema of the MIRO signals (circles in panel \textbf{c}) illustrating the asymmetry of MIRO between the CR-active and CR-passive polarities of $B$ for $\eta=1$ (red circles) and $\eta=-1$ (blue circles). The solid curve shows the  ratio $A^\text{D}(B)/A^\text{D}(-B)$ of the Drude absorption calculated for $\eta=1$ using parameters extracted from the dark transport measurements and transmission $T(B)$ shown in panel (b). 
		} \label{fig1}
	\end{figure*}
	
	Despite much progress in unified understanding of the above phenomena, the developed theory of MIRO is considered inadequate in view of reported puzzling insensitivity of MIRO to helicity of the incoming radiation \cite{smet:2005,herrmann:2016}. The microscopic theory \cite{dmitriev:2012} predicts a strong asymmetry of MIRO with respect to the polarity of the magnetic field $B$, applied perpendicular to the 2DES plane, in the case of circular polarization of  incident radiation. According to this theory, at low radiation intensity $I$ the photoresistance, $\delta R(B)=I A^\text{D}(B) f(|B|)$, can be represented as a product of the quasiclassical Drude absorptance, $A^\text{D}(B)$, which should be strongly enhanced near the cyclotron resonance (CR) either at positive or negative $B=\eta B_\text{CR}$ depending on the helicity $\eta=\pm 1$ of the incoming wave, and a mechanism-dependent function $f(|B|)$ which describes the magnetooscillations and is insensitive to the sign of $B$ or $\eta$. In sharp contrast to the expected pronounced $B$-asymmetry, 
	\begin{equation}
	\label{ratio}
	\delta R(B)/\delta R(-B)=A^\text{D}(B)/A^\text{D}(-B),
	\end{equation}
	the experiments revealed completely symmetric \cite{smet:2005} or weakly asymmetric \cite{herrmann:2016} magnetooscillations in the photoresistance. At the same time, these experiments demonstrated a strongly asymmetric transmittance $T(B)$ through the 2DES, with a single dip at $B=\eta B_\text{CR}$, thus apparently confirming the validity of the quasiclassical theory behind interrelated $A(B)$ and $T(B)$, as well as the purity of circular polarization of the incoming wave.
	
	Here we provide a counterexample of GaAs-based 2DES that exhibits an unmitigated helicity dependence in both transmittance and MIRO in full agreement with Eq.~(\ref{ratio}). We thus prove that, in contrary to a widespread view, the long debated polarization immunity is not a generic property of MIRO in solid-state 2DES but rather reflects certain yet unknown peculiarities of technological design and corresponding realization of disorder in particular structures. We also establish the necessary conditions to avoid extrinsic electrodynamic effects capable to produce an apparent polarization immunity of MIRO.

\begin{figure}
		\includegraphics[width=1\columnwidth]{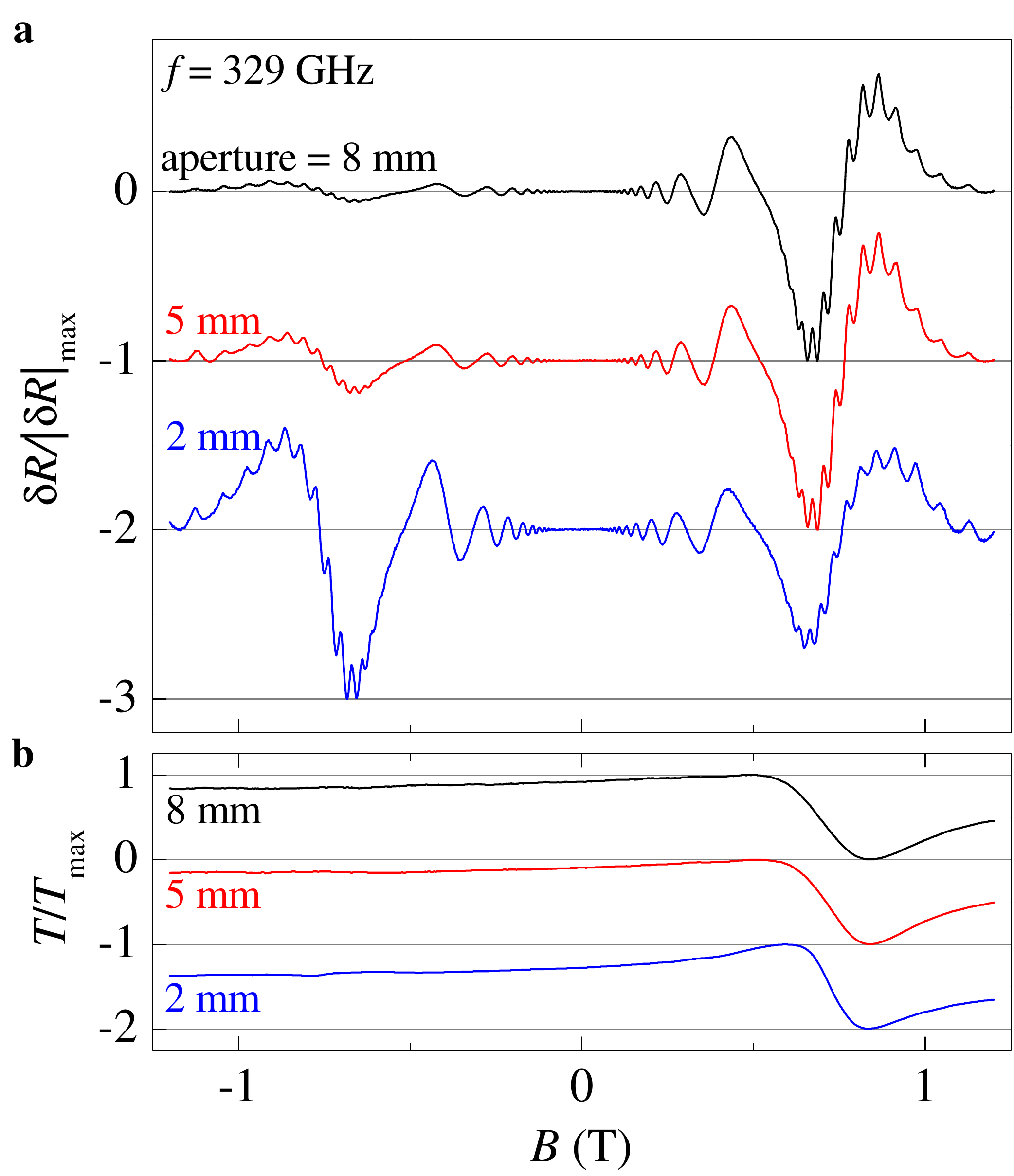}
		\caption{
			\textbf{a}, MIRO signal measured at $\nu = 329$\,GHz ($\lambda \approx 0.9$\,mm) using metallic apertures with different diameter as indicated. 
			\textbf{b}, Normalized transmission through the sample measured \textit{in situ}. }
		\label{fig2}
	\end{figure}

	{\it Methods.--}
	Our simultaneous transmittance and photoresistance measurements,  see Fig.~\ref{fig1}\,(a), were carried out on a heterostructure containing 2DES in a selectively doped 16-nm  GaAs quantum well with AlAs/GaAs superlattice barriers grown by molecular beam epitaxy~\cite{SM1,baba:1983,Friedland1996,umansky:2009,manfra:2014}. The lateral size of the van der Pauw sample was $10\times10\,$mm$^2$, with ohmic Ge/Au/Ni/Au contacts at the corners. The electron density and mobility extracted from magnetotransport measurements were $n = 7.0\times10^{11}\,$cm$^{-2}$ and  $\mu = 1.0\times 10^{6}\,$cm$^{2}$/Vs. The sample was irradiated from the substrate side through circular metallic (Fig.~\ref{fig2}) or absorber (Fig.~\ref{fig1}) apertures of varying diameter. Backward-wave oscillators (BWO) were used as stable sources of normally incident continuous monochromatic radiation with frequency $f = \omega/2\pi$ in the range between 50 and $500$\,GHz. A split-coil superconducting magnet provided a magnetic field oriented perpendicular to the sample surface. The photoresistance $\delta R$ (the difference of the resistance signals in the presence and absence of irradiation) was measured using the double-modulation technique~\cite{SM1}. In parallel with the photoresistance, the transmittance through the sample was measured using a liquid He-cooled bolometer. All presented results were obtained at temperature of $3.7$\,K.

	{\it Helicity dependence of MIRO, transmittance, and absorptance.--} 
	Figure~\ref{fig1} shows representative examples of simultaneously measured transmittance $T(B)$ and photoresistance $\delta R(B)$, recorded for both helicities $\eta=\pm 1$ of $f=\omega/2\pi=324$ GHz circularly polarized radiation. The transmittance traces in panel (b) display strong dips at the CRs, $B=\eta B_\text{CR}$, marked by arrows.
	The absence of any features at opposite $B = -\eta B_\text{CR}$ confirms a high degree of the circular polarization in the transmitted signal. The photoresistance $\delta R(B)$ in Fig.~\ref{fig1}\,(c) shows pronounced MIRO governed by the ratio $\omega/\omega_\text{c}$ of the radiation and cyclotron frequencies. In sharp contrast to previous reports \cite{smet:2005,herrmann:2016}, our results reveal a high asymmetry of MIRO with respect to polarity of $B$ which inverts with the change of radiation helicity $\eta$. This is one of our central observations which unequivocally demonstrates that the long debated polarization immunity is not a generic property of MIRO in solid-state 2DES.
	
	Moreover, we find an excellent quantitative agreement between the shape of CR in measured $T(B)$ and the $B$-asymmetry of the MIRO signal, in full accordance with Eq.~(\ref{ratio}). This agreement is illustrated in Fig.~\ref{fig1}\,(d) where we plot ratios $\delta R(\eta B)/\delta R(-\eta B)$ of MIRO signal between CR-active and CR-passive polarities of $B$, for both helicities. Here we use values of $\delta R$ at the opposite-lying MIRO extrema marked by circles in panel~(c). It is seen that the asymmetry is strongly enhanced in the region of CR reaching values of $\sim 30$ for the main MIRO extrema around $\omega=\omega_\text{c}$. Solid and dashed lines show the corresponding ratio $A^\text{D}(B)/A^\text{D}(-B)$ of the Drude absorption calculated using parameters extracted from the shape of $T(B)$ in panel~(b), for 100\% right circular polarization and for 98\% to 2\% mixture of right and left polarizations, respectively. An almost perfect agreement with Eq.~(\ref{ratio}) demonstrates an ultimate sensitivity of MIRO to circular polarization. It is important to emphasize that this comparison, detailed below, does not assume any specific microscopic mechanism of MIRO. It rather establishes that polarization dependences of MIRO and of Drude absorption are the same.
	
	{\it Modelling of transmittance and absorptance.--} 
	As detailed in Supplemental Material \cite{SM1}, the shape of $T(B)$ can be well reproduced using the standard expression, $T(B)=4/|s_1(1+Z_0\sigma_\eta )+s_2|^2$, with the complex dynamic conductivity $\sigma_\eta=\sigma_{xx}+i\eta\sigma_{yx}$ of 2DES taken in the classical Drude form, $\sigma^\text{D}_\eta=e^2 n/(\mu^{-1}-i B_\text{CR}+i\eta B)$, and with two complex parameters $s_{1,2}=\cos\phi-i n_r^{\mp 1}\sin\phi$ describing the Fabry-P\'{e}rot interference due to multiple reflections in the GaAs substrate. Here, the interference phase $2\phi=4\pi w/\lambda_r$ is given by the ratio of the sample thickness $w$ and the wavelength $\lambda_r=c/f n_r$, $n_r$ is the refractive index of the substrate, and $Z_0\approx 377$~$\Omega$ is the impedance of free space. The resulting Drude transmittance $T^\text{D}$ and Drude absorptance, $A^\text{D}=Z_0 T^\text{D}\mathrm{Re}\,\sigma^\text{D}_\eta$, are given by
	\begin{equation}
	\label{TD}
	T^\text{D}(B)=|\alpha|^2\left|1-\dfrac{\beta}{\mu^{-1}+\beta - i B_\text{CR}+i\eta B}\right|^2,
	\end{equation}
	\begin{equation}
	\label{AD}
	A^\text{D}(B)=\dfrac{Z_0|\alpha|^2 e^2 n/ \mu}{|\mu^{-1}+\beta - i B_\text{CR}+i\eta B|^2}, 
	\end{equation}
	where $\alpha=2/(s_1+s_2)$ and $\beta=e n Z_0/(1+s_1/s_2)$. The shape of measured $T(B)$ in Fig.~\ref{fig1} is well reproduced by Eq.~(\ref{TD})  using $\mu^{-1}=0.01$\,T extracted from magnetotransport measurements, $B_\text{CR}=m_\text{CR}\omega/e=0.83$\,T corresponding to the cyclotron mass $m_\text{CR}=0.071 m_0$, and $\beta=(0.183 + i\, 0.127)$\,T \cite{SM1}. On top of smooth Drude transmittance, our high resolution measurements are able to resolve weak quantum $\omega/\omega_\text{c}$-oscillations studied in Ref.~[\onlinecite{Savchenko2020b}]. The extracted parameters are used in calculated ratios $A^\text{D}(B)/A^\text{D}(-B)$ in Fig.~\ref{fig1}(d), which establish the validity of Eq.~(\ref{ratio}).

	\textit{Aperture dependence of the polarization-sensitive measurements. --}
	While the strong helicity dependence is unambiguously demonstrated in Fig.~\ref{fig1}, our further measurements have shown that this result can be strongly affected by metallic aperture that is typically used to avoid illumination of the contacts and sample edges in this type of experiments. To explore this  issue we performed a detailed study of the effect of aperture diameter on the helicity dependence. Figure~\ref{fig2} shows the $B$-dependencies of the photoresistance (top panel) and transmittance (bottom panel)  measured at 329\,GHz ($\lambda \approx 0.9$\,mm) with varying  diameter of the aperture. Top panel demonstrates that the MIRO asymmetry strongly depends on the diameter of aperture when other experimental parameters are kept unchanged. We clearly see that the MIRO ratio $\delta R(B)/\delta R(-B)$ is the largest for 8\,mm aperture and tends towards unity for the diaphragms smaller than the estimated size of the focal spot of about 4\,mm. On the contrary, the transmittance signal in the bottom panel does not considerably change for different apertures.

	We explain such behavior as follows: The MIRO signal is sensitive to the polarization of the local electromagnetic field. If the aperture size is too small such that the radiation interacts with the metallic edges, then the local polarization in the nearby 2DES will be linear regardless of the incident degree of the circular polarization. This follows from the boundary conditions requiring that the $E$--field can only have a normal component at the metal surface~\cite{Landau1984}. Therefore, the radiation that reaches the metal aperture results in a distortion of the local polarization and, consequently, of the detected photoresistance signal reflecting the local absorption. 
	
	The appearance of local linearly-polarized electric field can be understood considering a circularly polarized wave irradiating a circular metallic waveguide, see Supplemental Material \cite{SM1} for details. The distribution of local circularly-rotating electric fields that fulfil the boundary conditions is given by:
	\begin{equation}\label{Seq_circ}
	E_{+} = -i E J_0(k_r r), \qquad 
	E_{-} = -i e^{2 i \varphi} E J_2(k_r r). 
	\end{equation}
	Here $J_n$ are the Bessel functions of $n$-th order, $k_r$ is the radial wavevector of the lowest propagating mode in a circular waveguide, $E_\pm$ are right- and left-rotating circular fields, $E$ is the amplitude of the incident electric field. 
	\begin{figure}
		\includegraphics[width=1\columnwidth]{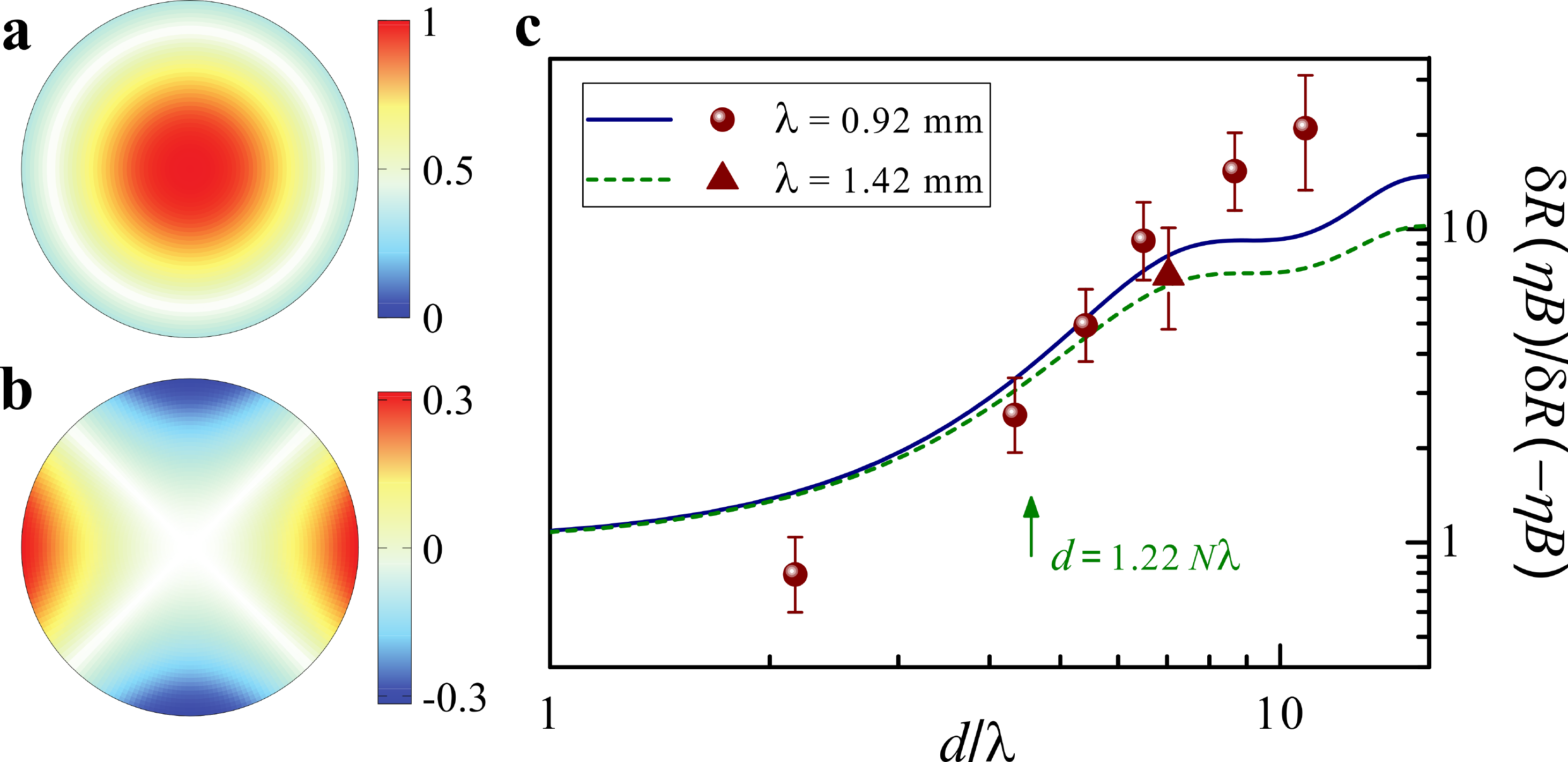} 
		\caption{
			\textbf{a}, \textbf{b}, Amplitudes of the circular polarization decomposition of the local field: the incident $E_+$ (\textbf{a}) and  distorted $E_-$ (\textbf{b}) polarizations.
			\textbf{c}, 
			MIRO ratio in experiments with varying diameter of the diaphragm. 
			Symbols represent the experiment, lines are calculated using Eq.\,(\ref{eq_focus}) and assuming that the local ratio of circular and linear polarizations is equal to the relative power transmitted through the aperture (see text for details). 
			Arrow indicates a characteristic size of the focal spot.
		}\label{fig_circ_amplitudes}
	\end{figure}
	The distributions of the circular components $E_{+}$ and $E_{-}$ are shown in Fig.~\ref{fig_circ_amplitudes}\,(a) and (b), respectively. 
	The amplitude of $E_{+} \sim J_0(k_r r)$ polarization is rotationally symmetric across the waveguide with a flat maximum in the middle. Obviously, this is the same polarization as that of the incident circularly-polarized beam. On the contrary, the $E_{-} \sim J_2(k_r r)$ is zero in the middle of the waveguide and changes sign four times along the wall of the waveguide.

	According to the Huygens-Fresnel principle the amplitude of the transmitted wave is proportional to the sum of secondary sources from all points on the wavefront. After integration of the $E_{-}$ fields, negative and positive regions cancel each other. Thereby, the distorted polarization is absent in the far field and produces no signal at the bolometer. That is the reason why all curves in Fig.~\ref{fig2}\,(b) show only one well-defined CR. In contrast to the transmittance, the absorbed power and MIRO are proportional to the square of the local near field. Therefore, the response to the distorted polarization does not vanish in resistivity signal.
	
	In Fig.\,\ref{fig_circ_amplitudes}\,(c) we show the ratio of MIRO amplitudes for active and passive circular polarizations as function of $d/\lambda$, where $d$ is the diameter of the metallic aperture and $\lambda$ is the radiation wavelength. 
	In order to estimate the ratio of intensities for active ($I_\text{A}$) and passive ($I_\text{P}$) circular polarizations we calculate the power ratio that is cut off by the aperture in the focus of the lens. 
	The radial distribution of the field in the focus\,\cite{born_book} is given by 
	\begin{equation} \label{eq_focus}
	I = I_0 \mathrm{J}_1(x)/x \ ,
	\end{equation}
	where $x=\pi d/2N\lambda$ is the normalized 
	aperture diameter, and $N=F/D\approx 3.6$ is the $f$-number of the focusing lens with diameter $D\approx 4.5$\,cm and focal length $F\approx 16$\,cm. Model curves in Fig.\,\ref{fig_circ_amplitudes}\,(c) are calculated using Eq.\,(\ref{eq_focus}) and taking $I_\text{A}/I_\text{P} = 50$ (solid line, $\lambda=0.92$\,mm) and $I_\text{A}/I_\text{P} = 32$ (dashed line, $\lambda=1.42$\,mm). The difference in the amplitudes originates from the frequency dependence of the MIRO ratio from Eq.\,(\ref{AD}), $I_\text{A}/I_\text{P} \sim f^2$. This yields a reasonable fit to the experimental points in spite of qualitative character of the arguments. Indeed, the MIRO ratio tends to unity for apertures below the diffraction spot $d=1.22 N \lambda$ indicated by the arrow.

	\textit{Discussion. --} 
	Our results demonstrate that the long-debated polarization immunity is not universal intrinsic property of MIRO in solid-state 2DES:
	(i) we detect strong intrinsic helicity dependence of MIRO that accurately follows the regular CR in the Drude absorption and 
	(ii) we show that extrinsic factors deteriorating the polarization state of radiation can be avoided using sufficiently large samples and apertures. On the other hand, very recent experiments \cite{Moench:2022}, where several samples of similar size were studied at terahertz frequencies, revealed a helicity immunity of the resonant photoresponse which directly measures the CR absorption via heating of 2DES. This clearly indicates that the helicity immunity problem remains and, moreover, has a more fundamental character, i.e. is not limited to MIRO. A possible reason \cite{Moench:2022} is that in particular 2DES the internal high-frequency response of electrons to a uniform external field can be essentially nonuniform, producing an intrinsic helicity distortion of the near field in the vicinity of individual strong scattering centers or inhomogeneities. Altogether, these results indicate an interesting research direction deserving further experimental and theoretical efforts.

	\textit{Conclusions. --}
	In the sub-terahertz frequency range we observe MIRO in GaAs/AlGaAs 2DES that are ultimately sensitive to the helicity of circularly polarized radiation. The helicity dependence accurately follows that of the Drude absorption, in full agreement with natural expectations. Moreover, we demonstrate that observation of intrinsic polarization dependence requires large sample sizes and large apertures. Otherwise, the near field reaching 2DES has a significant component of opposite helicity, which does not affect the transmittance measured in the far field, but essentially changes the asymmetry of the MIRO signal leading to apparent polarization immunity. Our results provide the opportunity to reliably test other systems potentially featuring an intriguing intrinsic polarization immunity of the photoresponse.
	
	\textit{Acknowledgements. --}
	The authors thank E. Mönch, D. A. Kozlov, D. G. Polyakov, and J. H. Smet for valuable discussions. We are grateful to A.\,K.\,Bakarov and A.\,A.\,Bykov for providing high-quality GaAs/AlGaAs samples. We acknowledge the financial support of the Austrian Science Funds (I 3456-N27,  I 5539-N), and of the German Research Foundation (Deutsche Forschungsgemeinschaft) via grant DM~1/5-1 (I.A.D.) and via sub-project A04 of the Project-ID 314695032 – SFB 1277 (S.D.G).
	
	\bibliography{MIRO_CR.bib}
	
	\pagebreak
	\widetext
	\begin{center}
		\textbf{SUPPLEMENTAL MATERIAL}
	\end{center}
	
	\setcounter{figure}{0}
	\setcounter{equation}{0}
	\renewcommand{\thesection}{S\arabic{section}}
	\renewcommand{\theequation} {S\arabic{equation}}
	\renewcommand{\thefigure} {S\arabic{figure}}
	\renewcommand{\thetable} {S\arabic{table}}
	
	\section{Helicity dependence of transmittance, Drude absorptance, and MIRO}
	\label{SM1}
	\noindent
	
	The main experimental finding of our work is strong helicity dependence of MIRO signal in the studied 2DES. It sharply contrasts all previous studies of solid-state 2DES which reported polarization immunity of MIRO in the case of circularly polarized radiation. Thus, our findings prove that the polarization immunity is not a universal intrinsic property of MIRO but rather a sample-dependent property possibly reflecting different disorder realization in particular structures. 
	
	Moreover, for our sample we are able to demonstrate that, in agreement with theoretical predictions \cite{dmitriev:2012}, the ratio of measured MIRO signal for opposite magnetic field polarities accurately follows that of the Drude absorptance $A^\text{D}(B)$, as long as extrinsic electrodynamic effects related to too small aperture or sample size in comparison to the laser beam and wavelength are suppressed, see Sec.~\ref{SM3}. As detailed below, this conclusion is based on analysis of simultaneously measured transmittance which readily provides the parameters of the corresponding $A^\text{D}(B)$ taking into account both strong reflection of radiation from 2DES in the region of CR and Fabry-P\'{e}rot interference in the dielectric substrate. For completeness, an explicit derivation of the corresponding standard expressions for transmittance and absorptance for uniform 2DES from Maxwell equations is provided in Sec.~\ref{SM2}. 
	
\begin{figure}
		\includegraphics[width=0.8\columnwidth]{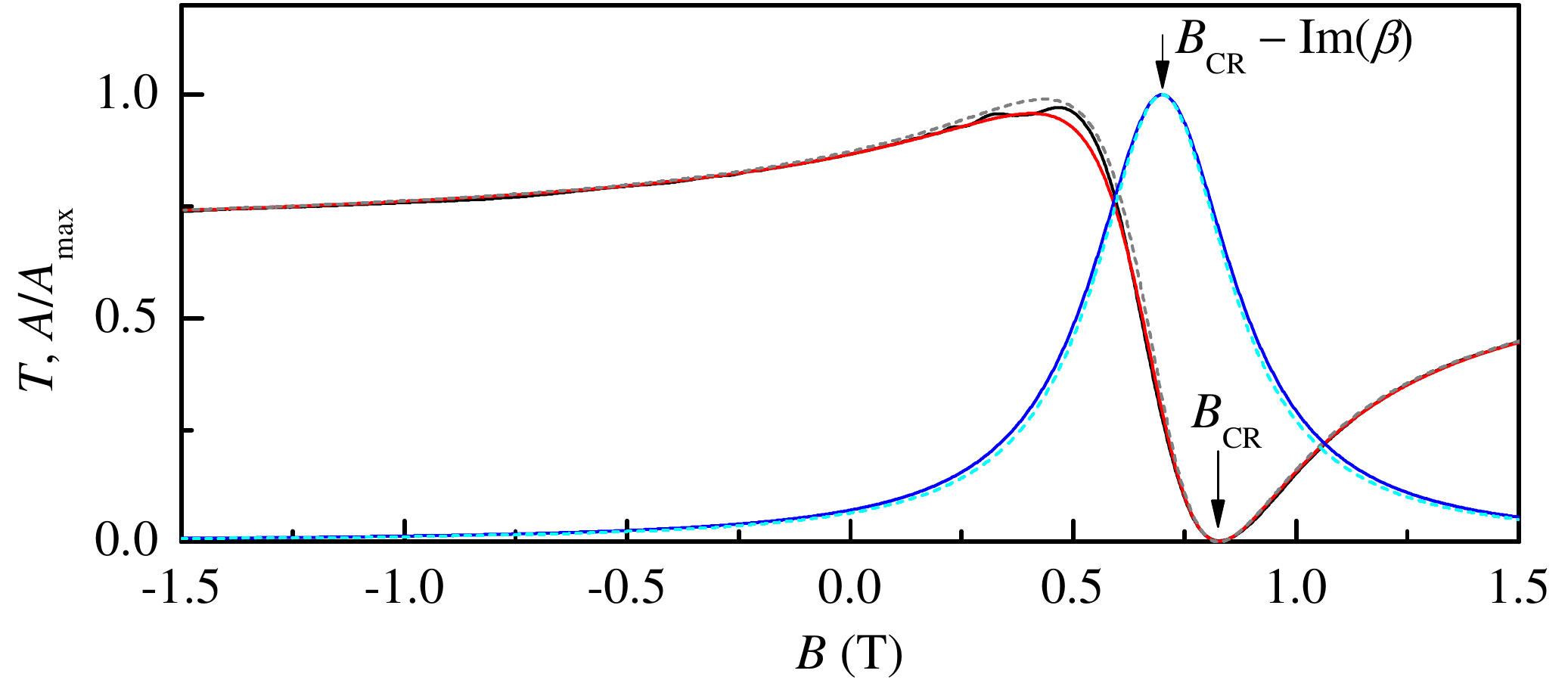}
		\caption{
			Magnetic field dependence of transmittance measured at frequency $f=324$\,GHz (black), its Drude fit using Eq.~(\ref{sTD}) (red), and the corresponding Drude absorption (\ref{sAD}), normalized to its maximal value, $A_\text{max}$ (blue). The dotted lines show transmission and normalized absorption in a clean limit, i.e for transport mobility $\mu \rightarrow \infty$.
		}
		\label{Tfit}
	\end{figure}

	In Fig.~\ref{Tfit} we show the transmittance data presented in Fig.~1 of main text together with the Drude fit obtained using Eq.~(2) of the main text (shown below and derived in Sec.~\ref{SM2}):
	\begin{equation}
	\label{sTD}
	T^\text{D}(B)=|\alpha|^2\left|1-\dfrac{\beta}{\mu^{-1}+\beta - i B_\text{CR}+i\eta B}\right|^2.
	\end{equation}
	In the classical Drude approximation (\ref{sTD}), the shape of transmittance is fixed by the position of the CR, $B_\text{CR}=2\pi f m_\text{CR}/e=0.83$~T, which can be accurately determined from the position of the deep minimum in measured $T(B)$; by the mobility $\mu=10^6$~cm$^2$/m s, which was extracted from dark resistance measurements, see Sec.~\ref{sec_light}, and by complex fitting parameter $\beta=e n Z_0/(1+s_1/s_2)$, see the main text and  Sec.~\ref{SM2}, which provides a combined description of strong reflection of radiation from 2DES in the region of CR and Fabry-P\'{e}rot interference due to multiple reflections in the dielectric substrate. Fit of the measured transmittance using $\beta\approx(0.183 + i\, 0.127)$~T is shown by red solid line. It is seen that the shape of measured $T(B)$ (black line) is well captured by Eq.~(\ref{sTD}), apart from weak quantum oscillations coupled to harmonics of CR which were studied in our previous work \cite{Savchenko2020b} and are not directly relevant to the present discussion of helicity dependence of MIRO. A dashed line in Fig.~\ref{Tfit} shows $T^\text{D}(B)$ obtained using Eq.~(\ref{sTD}) where all parameters are kept the same apart from $\mu^{-1}$, that is set to zero. The dashed and solid red lines are very close to each other, which reflects the fact that in our high-mobility and high-density 2DES, the CR broadening is mainly due to strong reflection of radiation from 2DES and not due to impurity scattering, $\mathrm{Re}\,\beta\gg\mu^{-1}=0.01$~T.  
	
	The analysis of $T(B)$ readily gives the corresponding shape of the Drude absorptance, 
	\begin{equation}
	\label{sAD}
	A^\text{D}(B)=\dfrac{Z_0|\alpha|^2 e^2 n/ \mu}{|\mu^{-1}+\beta - i B_\text{CR}+i\eta B|^2},
	\end{equation}
	illustrated by blue line in Fig.~\ref{Tfit}. Similar to transmittance, the shape of $A^\text{D}(B)$ (normalized to its maximal value $A_\text{max}$ in Fig.~\ref{Tfit}) is almost insensitive to scattering. This is illustrated by dashed blue line showing $A^\text{D}(B)/A_\text{max}$ in the clean limit $\mu\to\infty$. The value $A_\text{max}\propto \mu^{-1}$ is, of course, sensitive to scattering, but this does not affect the helicity dependence of absorptance and MIRO and thus is irrelevant for our discussion. It is also worth mentioning that the Fabry-P\'{e}rot interference in the substrate, producing an asymmetric shape of $T(B)$ in the case of $\mathrm{Im}\,\beta\neq 0$, also leads to a significant shift of the CR maximum in absorption from $B_\text{CR}$ to $B_\text{CR}-\mathrm{Im}\,\beta$, see arrows in Fig.~\ref{Tfit}.

	According to theoretical predictions \cite{dmitriev:2012}, MIRO in high-mobility 2DES in the case of circular polarization can be represented as a product
	\begin{equation}
	\label{sMIRO}
	\delta R(B)= I_\eta A_\eta^\text{D}(B) f(|B|),
	\end{equation}
	where $I_\eta$ and $\eta=\pm 1$ are intensity and helicity of the incident wave, $A_\eta^\text{D}(B)$ is the corresponding Drude absorption (\ref{sAD}), and $f(|B|)$ is a mechanism-specific function which describes magnetooscillations and depends only on the magnitude but not polarity of the magnetic field applied at normal to the 2DES plane. From the theoretical viewpoint, the relation (\ref{sMIRO}) in high-mobility 2DES remains valid as long as the intensity is low enough, such that the photoresponce remains linear, $\delta R(B)\propto I_\eta$. This linearity in our study is confirmed by measurements of MIRO at different power levels of the source, see Sec.~\ref{sec_diff_power}. The form of MIRO in Eq.~(\ref{sMIRO}) implies that the $B$-asymmetry (and, thus, the helicity dependence) of the Drude absorption and MIRO should be the same:
	\begin{equation}
	\label{sRatio}
	\delta R(B)/\delta R(-B)=A^\text{D}(B)/A^\text{D}(-B),
	\end{equation}
	see Eq.~(1) of the main text. This relation does not involve the mechanism-specific function $f(|B|)$ and thus can be verified independently, without full analysis of the detected MIRO within specific mechanisms which will be presented elsewhere.
	
	Comparison of the MIRO and Drude absorption ratios in Fig.~1(d) demonstrates that the relation (\ref{sRatio}) holds for our data. We attribute somewhat
	lower MIRO ratios in comparison to the Drude absorption ratios calculated for 100~\% $\eta=1$ circular polarization to the presence of a small component with an opposite helicity in the incoming radiation. To estimate the amount of such admixture, we also calculate the absorption ratio for an imperfect polarization,
	\begin{equation}
	\label{sRatioMixed}
	A^\text{D}(B)/A^\text{D}(-B)=\sum_\eta c_\eta A_\eta^\text{D}(B)/\sum_\eta c_\eta A_\eta^\text{D}(-B),
	\end{equation}
	and find that admixture of 2~\% of $\eta=-1$ circular polarization, i.e. $c_{-1}=0.02$ and $c_{1}=0.98$, is sufficient to perfectly fit the data for $\delta R(B)/\delta R(-B)$, see dashed line in Fig.~1(d).
	
\begin{figure}[h]
		\includegraphics[width=1\columnwidth]{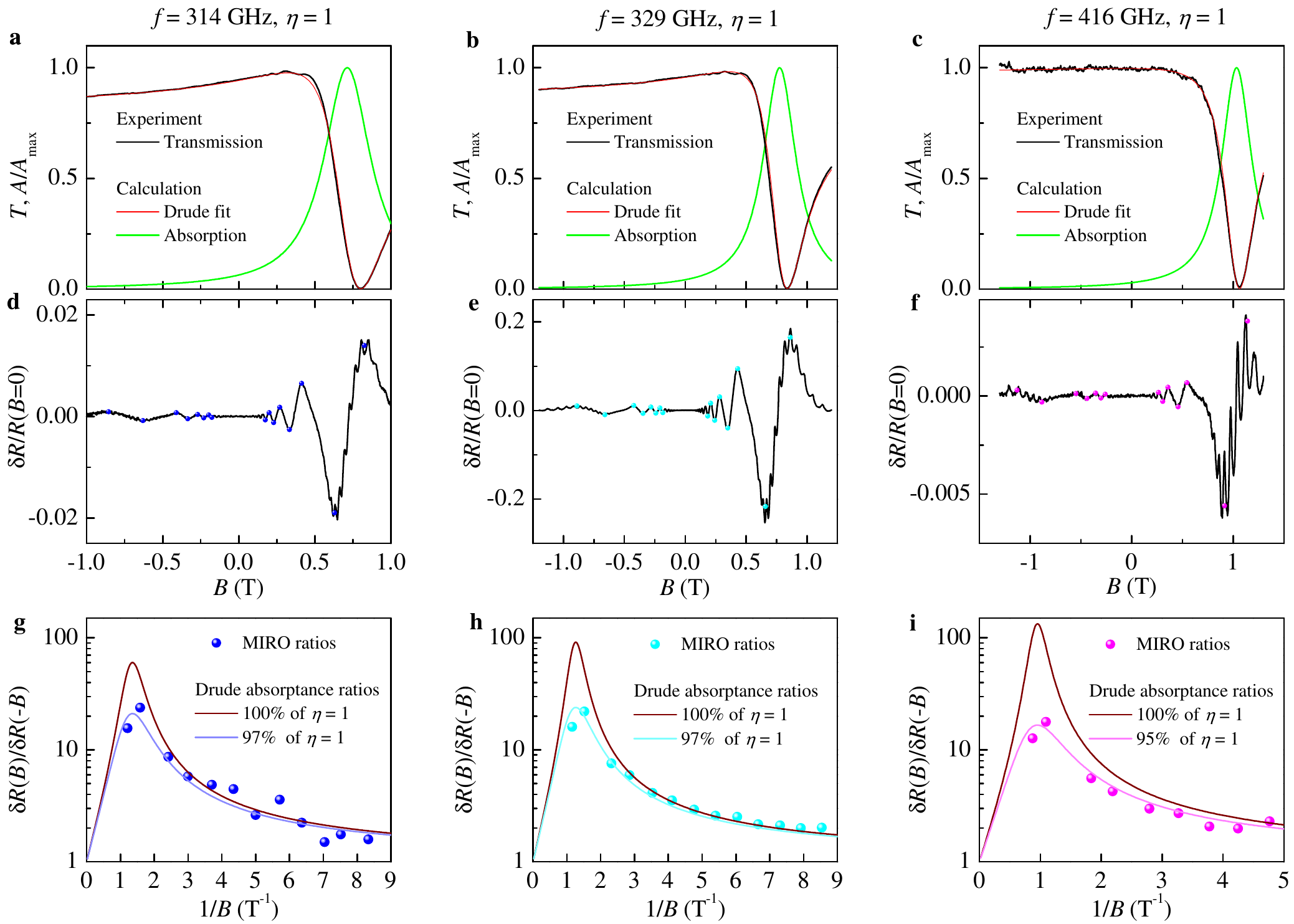}
		\caption{
			Experimental transmittance and MIRO measured at $f=314$\,GHz~(left), $329$\,GHz~(middle) and $416$\,GHz~(right), and their analysis.
			\textbf{a}, \textbf{b} and \textbf{c},~Magnetic field dependences of transmittance, their Drude fits and corresponding calculated normalized Drude absorption.
			\textbf{d}, \textbf{e} and \textbf{f},~Magnetic field dependences of MIRO in the longitudinal photoresistance $\delta R$, normalized to the zero field dark resistance $R(B=0)$. Dots mark MIRO extrema; the corresponding ratios $\delta R(B)/\delta R(-B)$ vs. $1/B$ are plotted in 
			\textbf{g}, \textbf{h} and \textbf{i} illustrating the asymmetry of MIRO between the CR-active and CR-passive polarities of $B$. 
			The wine curves show the ratio $A^\text{D}(B)/A^\text{D}(-B)$ of the Drude absorption calculated for 100\% $\eta=1$ polarization using parameters extracted from the dark transport measurements and transmission $T(B)$ shown in top panels. The experimental points are best fitted if the absorption is calculated using an admixture of 3\% (\textbf{g} and \textbf{h}) or 5\% (\textbf{i}) of $\eta=-1$ CR-passive component in the intensity of incident wave.
		}
		\label{fig_diam8}
	\end{figure}

	The data in Fig.~1 were obtained using the absorbing aperture of $\approx 10$~mm diameter at $f=324$~Ghz. Figure \ref{fig_diam8} shows several further examples of the analysis presented above, but now for data obtained with metallic aperture of $8$~mm diameter at three different frequencies of $\eta=1$ radiation. Also here we observe an excellent agreement between helicity dependence ($B$-asymmetry) of MIRO and absorption. The best fit in the region of CR is obtained for using an admixture of 3\% (for two lower $f$) or 5\% (for the highest $f$) of $\eta=-1$ component in the intensity of incident wave. Apart from confirmation of the results in Fig.~1, it shows that using $8$~mm metallic aperture at $f$ above $300$ GHz is sufficient to suppress the extrinsic effects capable to mask strong intrinsic helicity dependence of MIRO. These effects, see Sec.~\ref{SM3}, become strong at lower frequencies, see Sec.~\ref{sec_tr}, or for smaller aperture size, see Fig. 2 of the main text. Thus, our results, establishing the conditions necessary to avoid deterioration of the polarization state of radiation reaching 2DES, provide the opportunity to reliably test other systems potentially featuring an intriguing intrinsic polarization immunity of the photoresponse.

	\newpage
	\section{Transmittance, MIRO, and MIRO ratios at different frequencies}
	\label{SM2a}

	\newpage
	\section{Derivation of expression for transmission of a plane electromagnetic wave though a sample containing a uniform isotropic 2DES}
	\label{SM2}
	
	\noindent
	Here we apply a standard theory of transmission though a stratified media to our case of a circularly polarised wave normally incident on a sample
	containing a uniform isotropic 2DES, and obtain expressions for transmittance and absorptance used in the main text and Sec.~\ref{SM1}.
	
	Consider an incident plane electromagnetic wave with electric field given by the real part of ${\bf E}_i= E_0 {\mathbf e}_\eta \exp(-i\omega t +i k z)$ where the polarization vector ${\bf e}_\eta={\bf e}_x +i\eta{\bf e}_y$ corresponds to a circular polarization with helicity $\eta=\pm 1$. The wave is normally incident on the back (substrate) side of the sample at $z=0$. The sample is modelled as a dielectric slab with refractive index $n_r$ and thickness $w$ that contains a 2DES at a negligibly small distance $w-z_1\ll 1/k$ from the opposite front interface at $z=w$. The wave is partially transmitted and partially reflected at both back and front interfaces. By continuity of the tangential electric field at all interfaces (note that normal components of the high-frequency field do not appear in this problem in the case of a uniform 2DES), the electric field acting on electrons in this setup is given by ${\bf E}|_{z=w}=t{\bf E}_i|_{z=0}$, and the same transmission amplitude $t$ determines the transmittance $T(B)=|t|^2$. For a uniform isotropic 2DES, characterised by a local dynamic conductivity tensor $\hat\sigma$ with components $\sigma_{xx}=\sigma_{yy}$ and $\sigma_{xy}=-\sigma_{yx}$, the sense of rotation of fields in all reflected and transmitted partial waves is the same, and coincides with that of the induced electrical current in 2DES, ${\bf j}=\hat\sigma {\bf E}=\sigma_\eta {\bf E}$. Indeed, the conductivity tensor $\hat\sigma$ is diagonal in the helicity basis, $\hat\sigma{\bf e}_\eta=\sigma_\eta {\bf e}_\eta$ with $\sigma_\eta=\sigma_{xx}+i\eta\sigma_{xy}$. Therefore, propagation of both circularly polarized modes can be considered independently. The problem is reduced to determination of the scalar coefficients $c_\pm(z)$ in electric and magnetic field distributions in the whole space,
	\begin{eqnarray}\nonumber
	{\bf E}(z)&=&E_0 {\mathbf e}_\eta e^{-i\omega t}[c_+(z)e^{i k z}+c_-(z)e^{-i k z}],\\
	Z_0{\bf H}(z)&=&(k c/\omega) E_0 ({\mathbf e}_z\times{\mathbf e}_\eta) e^{-i\omega t}[c_+(z)e^{i k z}-c_-(z)e^{-i k z}],
	\end{eqnarray}
	where $Z_0$ is the vacuum impedance, and the wave vector $k=n_r\omega/c$ inside and $k=\omega/c$ outside the sample. According to the Maxwell equations, both ${\bf E}$ and ${\bf H}$ are continuous at the dielectric interfaces. Across the 2DES plane, the magnetic field experience a jump given by ${\bf H}(w+0)-{\bf H}(w-0)=-{\mathbf e}_z\times{\bf j}$. Implementing these boundary conditions, one arrives at a matrix equation $A(w+0)={\cal M}_\text{2DES}{\cal M}_\text{diel}A(0)$ for a vector $A(z)=(E, Z_0 H)^T$ constructed from the electric and magnetic field amplitudes. This equation,
	explicitly given by
	\begin{equation}
	\label{transfer}
	\binom{t}{t}=
	\begin{pmatrix}
	1     & 0\\
	-Z_0 \sigma_\eta & 1
	\end{pmatrix}
	\begin{pmatrix}
	\cos\phi     & i n_r^{-1}\sin\phi\\
	i n_r \sin\phi & \cos\phi 
	\end{pmatrix}
	\binom{1+r}{1-r}\,,\qquad \phi=w n_r\omega/c,
	\end{equation}
	relates the value of $A(w+0)=(t, t)^T$ for the wave transmitted through the front interface to $A(0)=(1+r, 1-r)^T$ combining the incident and reflected waves at the back interface. Two transfer matrices ${\cal M}_\text{diel}$ and ${\cal M}_\text{2DES}$ describe propagation through the dielectric substrate and 2DES, correspondingly. Solution of Eq.~(\ref{transfer}) yields the standard expression for transmittance, 
	\begin{equation}
	\label{T}
	T(B)=|t|^2=\dfrac{4}{|s_1(1+Z_0\sigma_\eta )+s_2|^2},\qquad s_{1,2}=\cos\phi-i n_r^{\mp 1}\sin\phi,
	\end{equation}
	as well as related expressions for reflectance $|r|^2$ and absorptance $A(B)=1-|r|^2-|t|^2=Z_0 |t|^2 \text{Re}\sigma_\eta$. 
	
	With  the complex dynamic conductivity $\sigma_\eta$ of 2DES taken in the classical Drude form, $\sigma^\text{D}_\eta=e^2 n/(\mu^{-1}-i B_\text{CR}+i\eta B)$, from Eq.~(\ref{T}) one obtains Eqs.~(2) and (3) of the main text for the Drude transmittance $T^\text{D}(B)$ and absorptance $A^\text{D}(B)$. In Sec.~\ref{SM1}, these expressions are used to demonstrate that the $B$-asymmetry of the Drude absorptance and MIRO in our sample are the same, as long as extrinsic electrodynamic effects are excluded, see Sec.~\ref{SM3}. In Sec.~\ref{SM1}, it is also demonstrated that the shape of both  $T^\text{D}(B)$ and $A^\text{D}(B)$ in our sample are almost insensitive to the impurity scattering, and can well reproduced using conductivity $\sigma^\text{C}_\eta=i e^2 n/(B_\text{CR}-\eta B)$ of a clean 2DES, see dashed lines in Fig.~\ref{Tfit}.

	\newpage
	\section{Influence of extrinsic factors which explains differences in polarization dependence of transmission and photoresistance}
	\label{SM3}
	
	Mathematically, the appearance of local linearly-polarized electric field can be understood considering a circularly polarized wave falling on a circular metallic waveguide. Although in the general case several eigenmodes of the waveguide will be excited, the mode closest to a circular plane wave is the TE$_{1,1}$ mode. Electric fields of this mode in cylindrical coordinates are given by (common factor is $\exp(\imath (n \varphi + k_z z - \omega t))$):
	\begin{eqnarray}
	\nonumber E_r&=&- i E_0 \frac{J_n(k_r r)}{k_r r} \, , \\
	\label{eq_te11} E_{\varphi}&=&E_0 \frac{dJ_n(k_r r)}{d(k_r r)} \, ,\\
	\nonumber E_z&=&0 \, .
	\end{eqnarray}
	Here $J_n$ are the Bessel functions of the $n$-th order, $n = 1$ for TE$_{1,1}$ mode, the radial and axial wavevectors are connected by the dispersion relation $\omega^2 / c^2 = k_r^2 + k_z^2$ and the radial wavevector is
	given by $k_r = u / R$ with the first root of the derivative of the first Bessel function $u \approx 1.8412$.
	
\begin{figure*}[h]
		\begin{minipage}{0.32\columnwidth}
			\centering
			\includegraphics[width=0.65\columnwidth, clip]
			{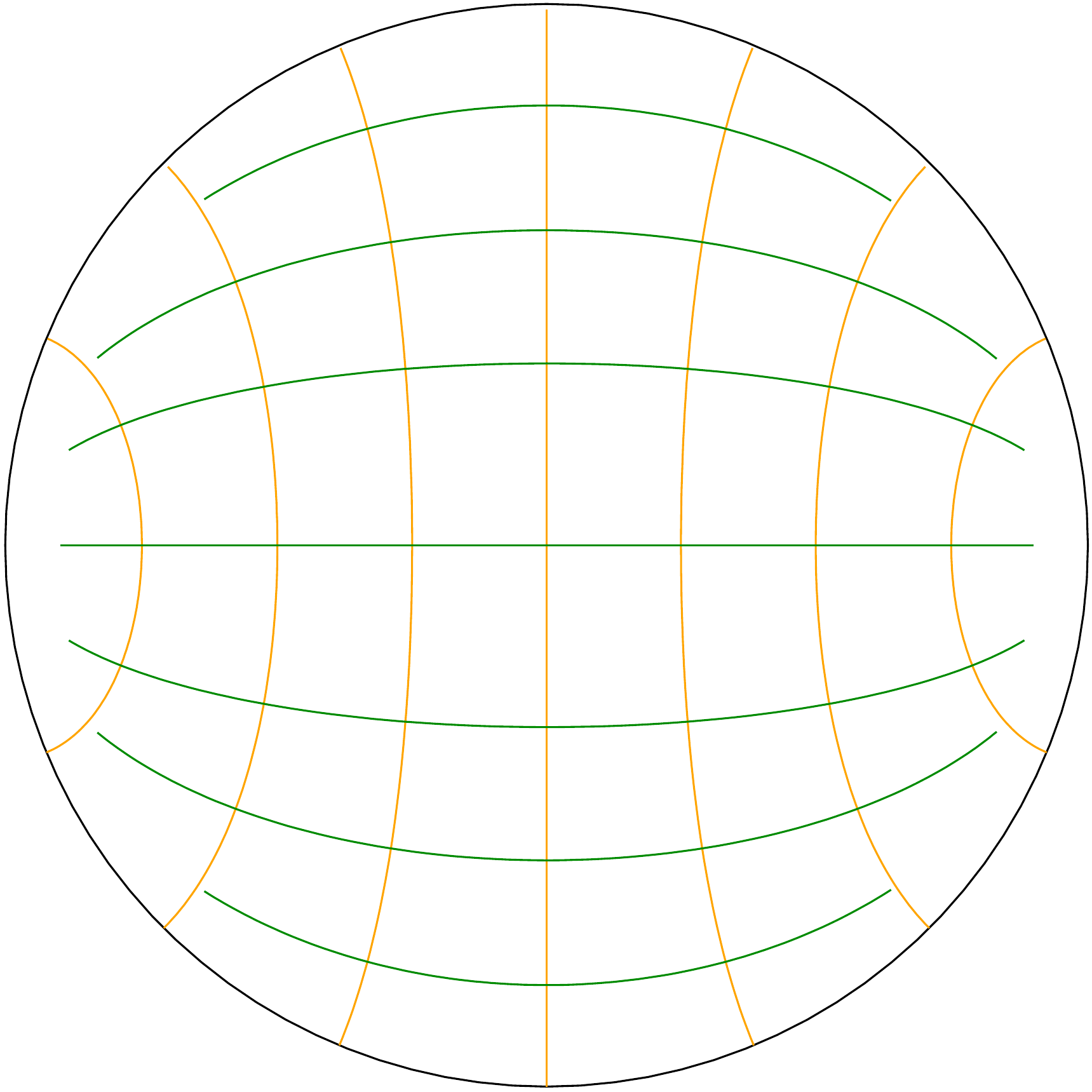}
		\end{minipage}
		\begin{minipage}{0.32\columnwidth}
			\centering
			\includegraphics[width=0.9\columnwidth, clip]
			{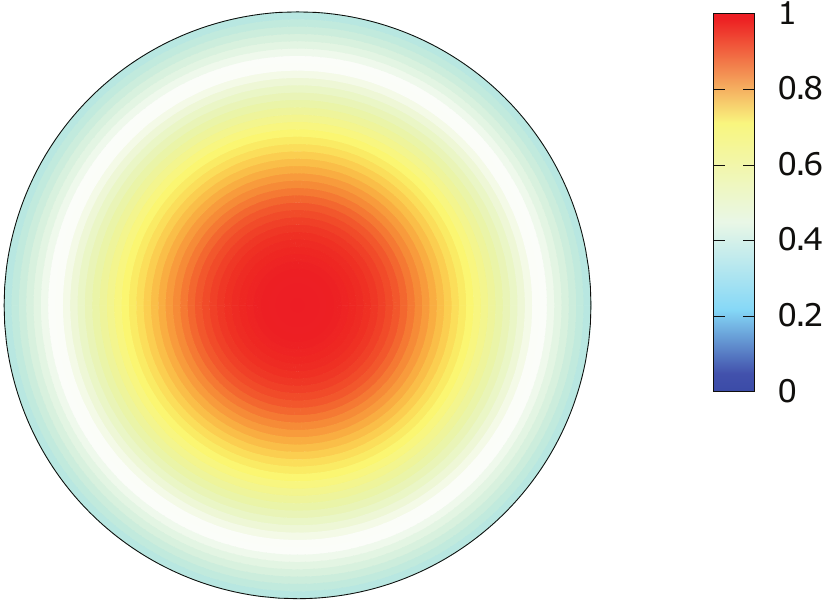}
		\end{minipage}
		\begin{minipage}{0.32\columnwidth}
			\centering
			\includegraphics[width=0.9\columnwidth, clip]
			{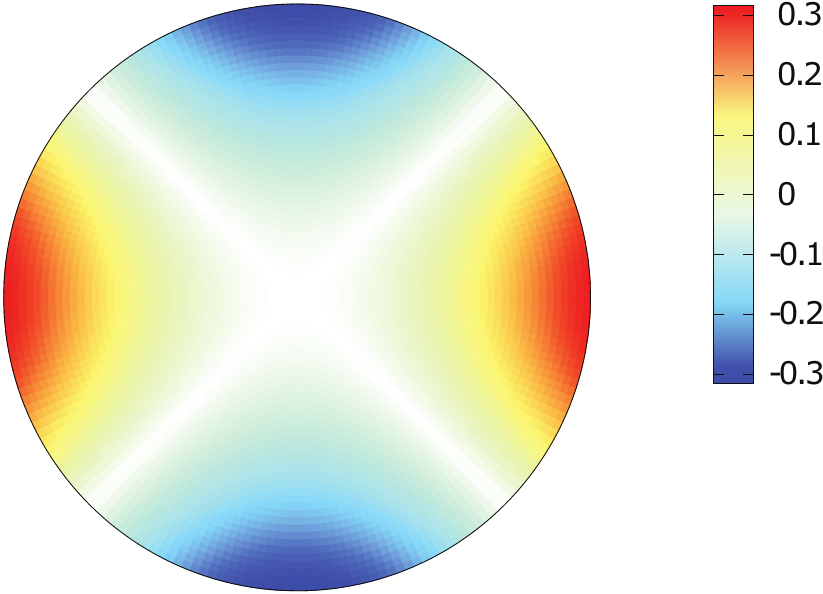}
		\end{minipage}
		\caption{
			Left panel: momentarily distribution of electric (orange) and magnetic (green) fields for a TE$_{1,1}$ mode of a circular waveguide. Middle and right panels: decomposition of the eigenmode into local right- (middle) and left-rotating electric fields.
		}\label{sfig_circ_amplitudes}
	\end{figure*}

	The calculated field lines of  electric (orange lines) and magnetic (green lines)
	fields of TE$_{1,1}$ mode are shown in the left panel of
	Fig.\,\ref{sfig_circ_amplitudes}. The electric field
	lines start and end at the metallic surfaces and are perpendicular to the
	cylinder axis. This field distribution rotates clockwise  as a function of time.
	In the center of the waveguide the TE$_{1,1}$ mode reproduces exactly the circularly polarized wave and fulfils the boundary conditions of a normal electric field at the edges. Locally, electric fields can be decomposed into a sum of left-rotation and right-rotating circular fields via
	$  E_\pm =E_x \pm i E_y $. 
	Here $E_x$ and $E_y$ are the fields in Eq.\,(\ref{eq_te11}) expressed in the cartesian coordinates. Finally, the momentarily distribution of local circularly-rotating fields  are given by:
	\begin{equation}\label{eq_circ}
	E_{+} = -\imath e^{2 \imath \varphi} E J_2(k_r r), \qquad
	E_{-} = -\imath E J_0(k_r r).
	\end{equation}

	The distributions of the circular components $E_{-}$ and $E_{+}$
	are shown in Fig.~\ref{sfig_circ_amplitudes} in the middle and right panels, respectively. The amplitude of $E_{-}$ polarization is rotation-symmetric across the waveguide with a flat maximum	in the middle. Obviously, this is the same polarization as that of incident
	circularly-polarized beam. On the contrary, the $E_{+}$ polarization is zero in the middle of the waveguide and changes sign four times along the wall of the waveguide (see red and blue regions in the right panel). This is the distorted polarization due to the boundary condition at the metal surface.
	
	Importantly, the distorted polarization is absent in the far field and produces zero signal at the bolometer. According to the Huygens-Fresnel principle the amplitude of the transmitted wave is proportional to the sum of secondary sources from all of the points on the wavefront. After integration of the $E_{+}$ fields, negative and positive regions cancel each other.	Contrary to a linear response in transmittance, the MIRO signal reflects local near field containing both circular components. Therefore, the response to the distorted polarization does not vanish in MIRO signal.
	
	\begin{figure}[tbp]
		\includegraphics[width=0.5\columnwidth]{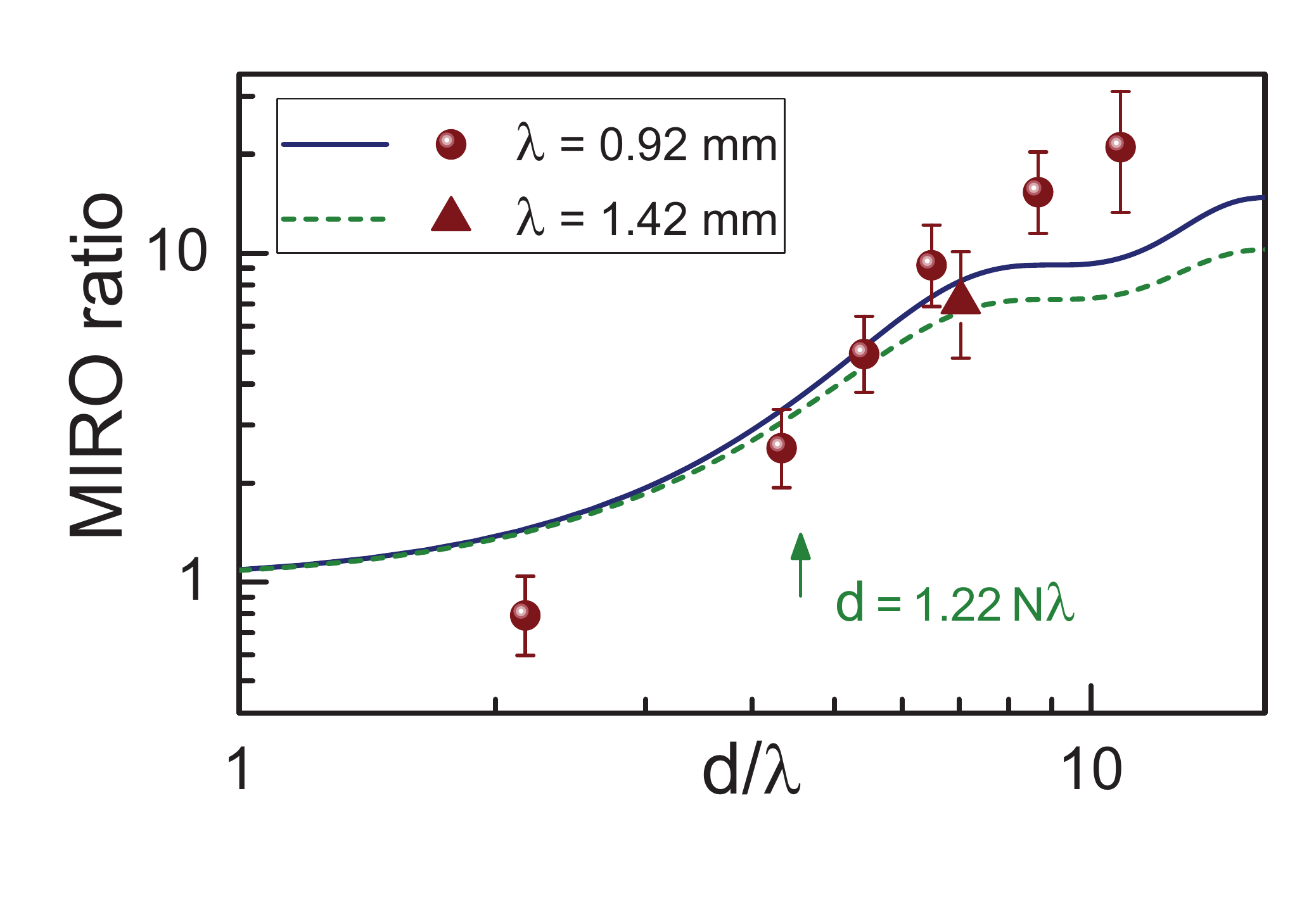}
		\caption{\textit{Dependence of the MIRO ratio on the sample aperture.}
			MIRO ratio in experiments with varying diameter of the diaphragm. Symbols - experiment, lines are calculated using Eq.\,(\ref{Seq_focus}) and assuming that the the local ratio of circular and linear polarizations is equal to the relative power transmitted through the diaphragm (see text for details). Arrow indicates a characteristic size of the focal spot.}
		\label{fig_diam}
	\end{figure}
	
	Solid symbols in Fig.\,\ref{fig_diam} show the ratio of MIRO amplitudes for active and passive circular polarizations as function of the size of the aperture. To compare the results in different experiments we plot MIRO as a function of ratio $d/\lambda$, where $d$ is the diameter of the metallic diaphragm and $\lambda$ is the wavelength. The data have been obtained for $\nu = 325$\,GHz (spheres) and $\nu = 211$\,GHz (triangles).
	Although an estimate of the local field distributions based on the TE$_{1,1}$ mode provides a qualitative explanation of the observed effect, it does explain the relation between MIRO ratio and the diameter of the aperture.
	In order to get an  estimate of this relation we suggest to attribute it to a power ratio that is cut off by the diaphragm in the focus of the lens. The radial distribution of the field in the focus\,\cite{born_book} is given by 
	\begin{equation} \label{Seq_focus}
	I = I_0 \mathrm{J}_1(x)/x \ ,
	\end{equation}
	where $x=\pi d/2N\lambda$ is the renormalized ratio of the diaphragm diameter $d$ and the wavelength $\lambda$. Here $N=F/D\approx 3.6$ is the $f$-number of the focusing lens with diameter $D\approx 4.5$\,cm and focal length $F\approx 16$\,cm. We assume that the ratio of the circularly- and linearly-polarized light in the sample plane is equal to the fraction of the power transmitted through the diaphragm. Model curves in Fig.\,\ref{fig_diam} are calculated using Eq.\,(\ref{Seq_focus}) and taking  $I_A/I_\text{P} = 50$ (solid line) and $I_\text{A}/I_\text{P} = 32$ (dashed line)  for $\nu = 325$\,GHz  and $\nu = 211$\,GHz, respectively. The difference in the amplitudes originates from the frequency dependence of the MIRO ratio~\cite{herrmann:2016} $I_A/I_\text{P} \sim \nu^2$. We conclude a reasonable fit to the experimental points in spite of qualitative character of the given arguments.

	\newpage
	\section{Power dependence of MIRO}
	\label{sec_diff_power}

	\begin{figure}[h]
		\includegraphics[width=0.95\columnwidth]{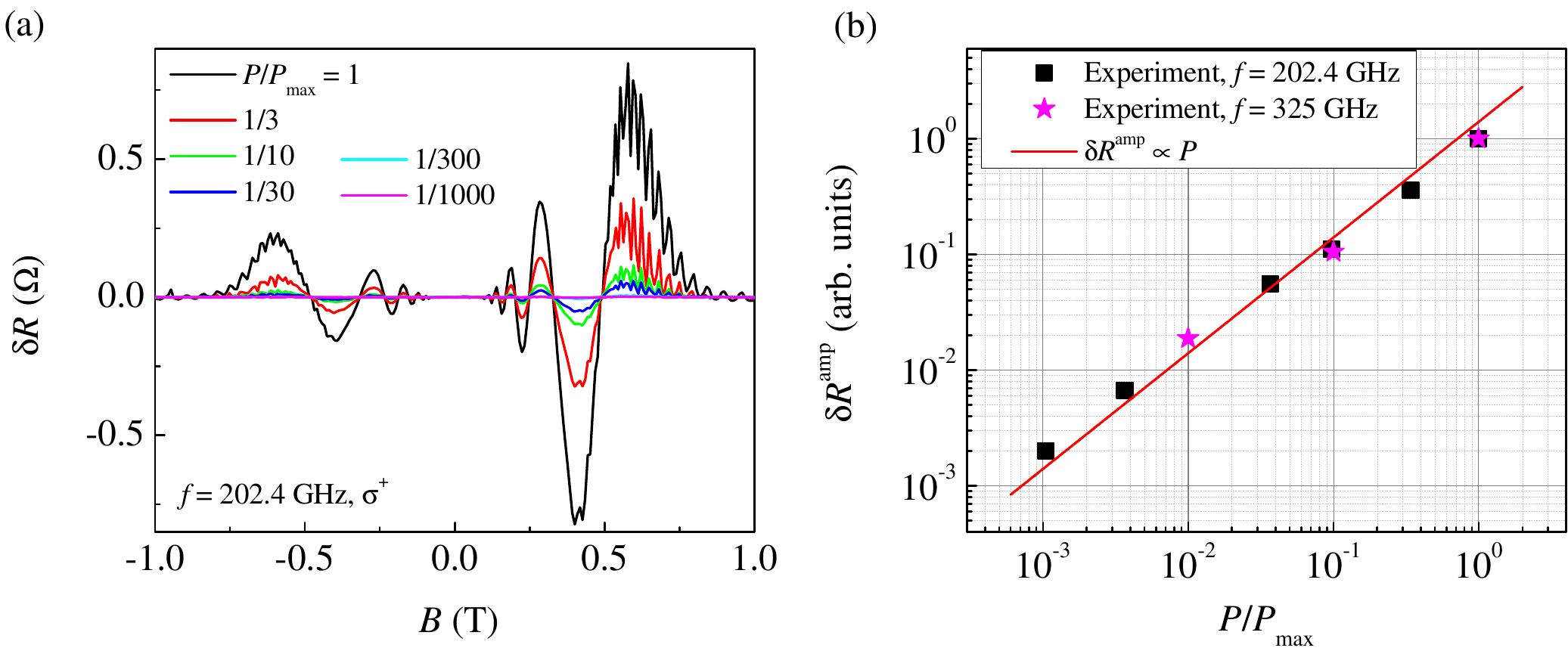}
		\caption{
			(a)~Magnetic field dependences of photoresistance $\delta R(B)$ measured at frequency $f = 202.4\,$GHz and for the right-hand circular polarization for different levels of the radiation power $P$ (in units of full power $P_\text{max}$, see legend).
			(b)~The amplitude of MIRO $\delta R^\text{amp}$ for $f = 202.4$ and 325\,GHz versus radiation power $P$ (symbols).   
			The straight line demonstrates that the MIRO amplitude is proportional to $P$ in the whole studied range.
		} \label{Diff_power}
	\end{figure}

	\newpage
	\section{Transmittance and MIRO at different frequencies}
	\label{sec_tr}
	
\begin{figure}[h]
		\includegraphics[width=1\columnwidth]{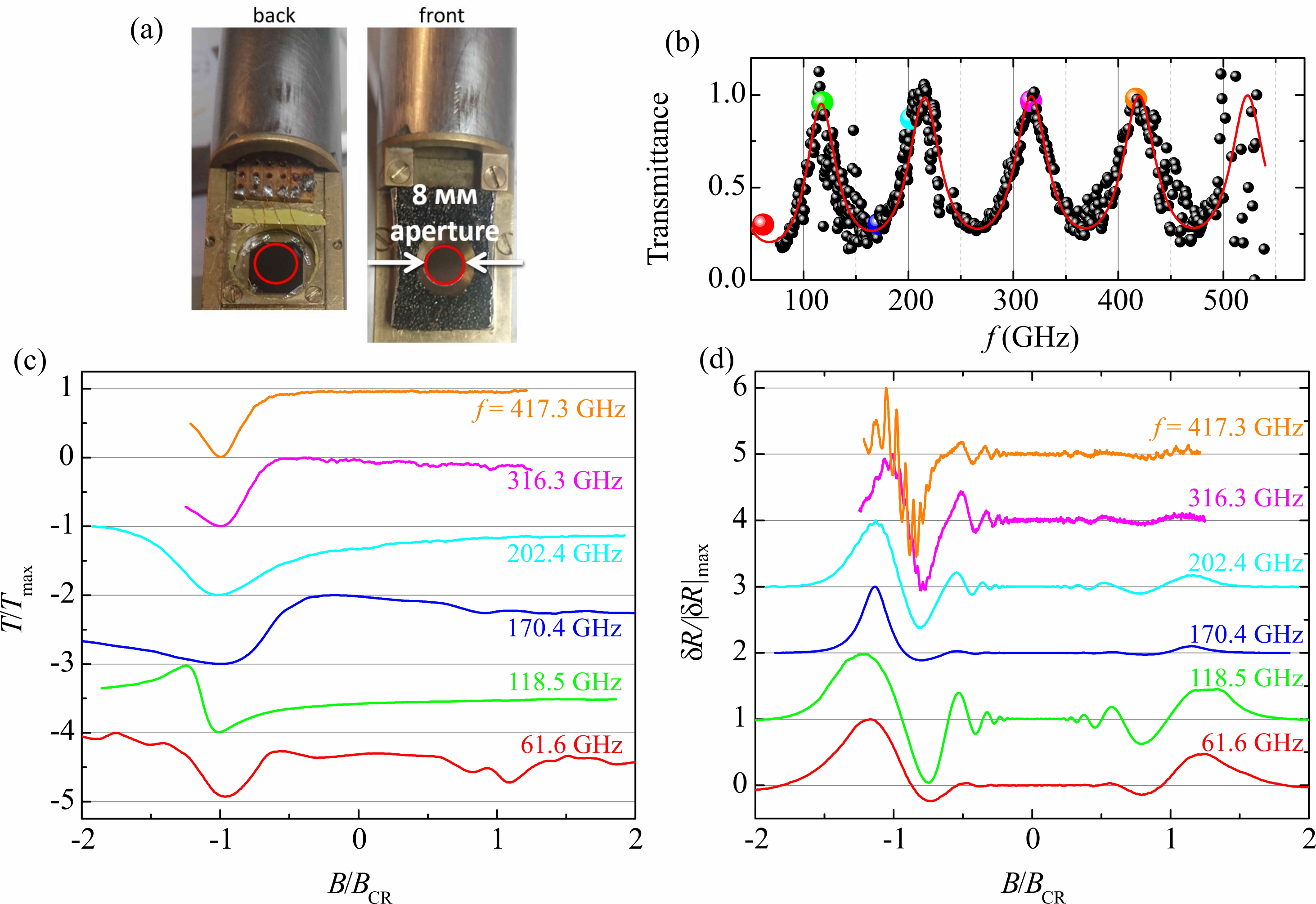}
		\caption{
			(a)~Mounting of the sample. 
			(b) Transmittance $T(f)$ at $B=0$ as a function of radiation frequency $f$.
			Symbols represent the experimental data points, while the red line shows a fit according to Eq.~(2) of the main text [Eq.~(\ref{T}) in SM]. Big color spheres correspond to frequencies at which measured transmittance and MIRO are presented on bottom panels. 
			(c) and (d)~Dependences of normalized to its maximum value transmittance $T/T_\text{max}$ and photoresistance $\delta R/| \delta R|^\text{max}$ vs $B/B_{CR}$ measured at indicating frequencies and for the left-hand circular polarization.
		} \label{T_vs_f_MIRO_diff_f}
	\end{figure}

	The sample insert consists of a fixed part with the beam aperture and the movable part with the sample. 
	The sample is fixed on a thin (6 to 13\,$\mu$m) Mylar foil which is clipped to the movable rod of the insert~(Fig.~\ref{T_vs_f_MIRO_diff_f}\,(a)). 
	
	In Fig.~\ref{T_vs_f_MIRO_diff_f}\,(b) we show the frequency dependence of transmittance $T(f)$ at $B=0$ which clearly demonstrates the Fabry-P\'{e}rot interference due to internal reflections between back and front interfaces of the sample.
	Black spheres represent the experimental data, while the red line is calculated according to Eq.~(2) of the main text (or Eq.~(\ref{T}) in SM). 
	Big color spheres correspond to 6 frequencies at which transmittance and MIRO are measured on panels~(c) and~(d). 
	
	In Fig.~\ref{T_vs_f_MIRO_diff_f}\,(c) and (d) we show the dependences of normalized transmittance and photoresistance on normalized $B/B_{CR}$ measured using 8\,mm aperture and circularly polarized radiation.
	Panel~(d) demonstrates that the MIRO asymmetry strongly depends on the frequency -- it increases as frequency goes up. 
	This observation can be explained by the influence of the metallic aperture at low ratios of the aperture size vs. the radiation wavelength. 
	This finding is in line with observation and explanation of the influence of the aperture size on MIRO asymmetry, see Fig.~2 and discussion in the main text and in Sec.~\ref{SM3}.

	\newpage
	\section{MIRO measurements using double modulation technique}
	
	Usually the microwave-induced resistance oscillations (MIRO) are measured under continuous microwave illumination. 
	The photoresponce is thereby obtained by comparing the magnetic field dependencies $R^\text{on}$ and $R^\text{off}$ of the longitudinal resistance in the presence and absence of radiation. 
	But at low radiation power the amplitude of MIRO becomes small, and this approach is not optimal. 
	Moreover, in our setup the transmittance of radiation is measured using a mechanical chopper such that the incoming radiation is modulated at frequency $f_\text{chopper}$. 
	Therefore, we measure photoresistance $\delta R = R^\text{on} - R^\text{off}$ using a double modulation technique specified below.
	
\begin{figure*}[b]
		\begin{minipage}{0.5\columnwidth}
			\centering
			\includegraphics[width=1\columnwidth, clip]{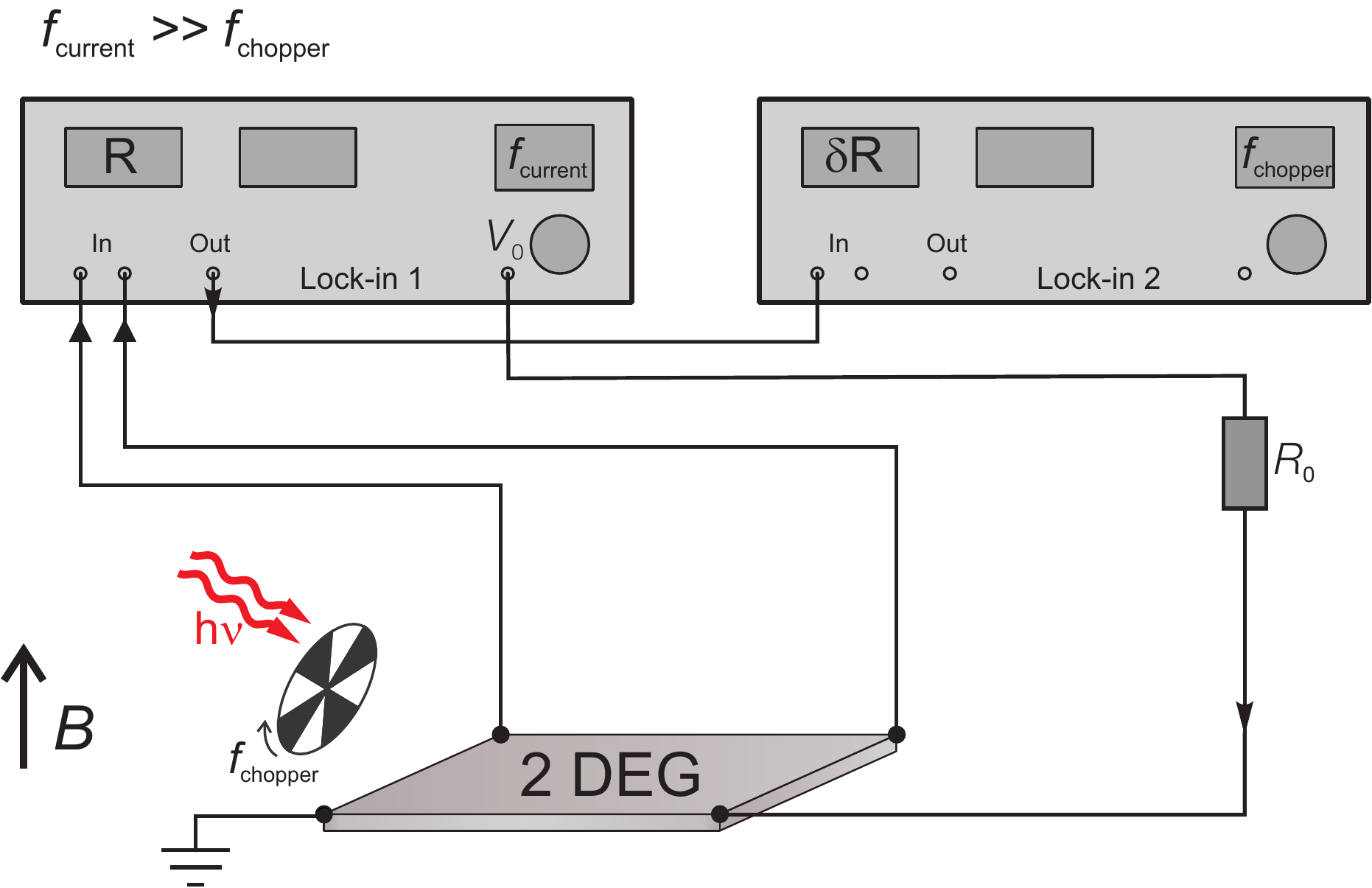}
		\end{minipage}
		\begin{minipage}{0.9\columnwidth}
			\centering
			\includegraphics[width=1\columnwidth, clip]{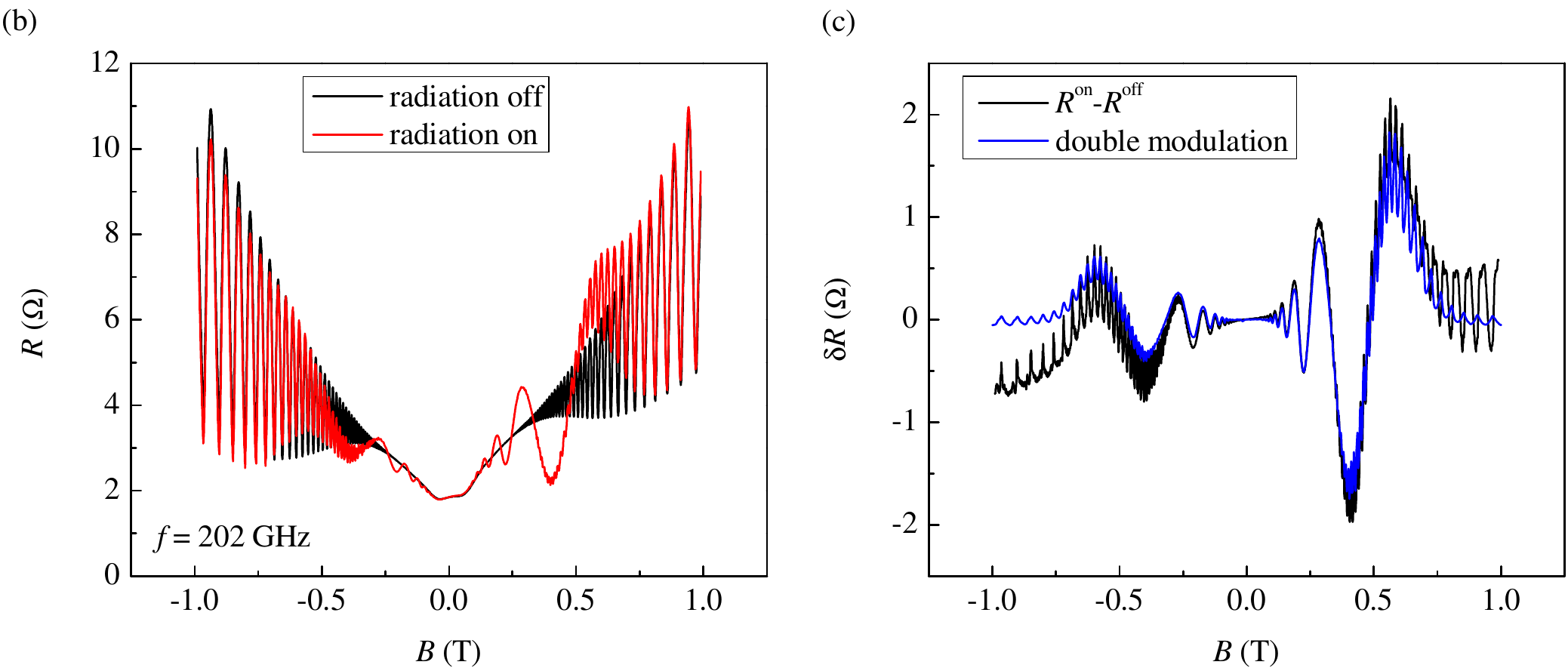}
		\end{minipage}
		\caption{
			(a)~Illustration of the double-modulation technique used in the measurements of photoresistance $\delta R$.
			Here the chopper frequency $f_\text{chopper} =23\,$Hz and the bias current is applied at higher frequency
			$f_\text{current} = 433\,$Hz. 
			The difference photoresistance signal $\delta R$ is obtained by using two lock-ins connected in series, where the first one collects the signal which oscillates in phase with the applied bias at frequency $f_\text{current} = 433\,$Hz, while the second collects the part $\delta R = R^\text{on} - R^\text{off}$ of the resulting dc resistance which oscillates in phase with the microwave power modulated at frequency $f_\text{chopper} =23\,$Hz.
			(b)~Direct measurement of the resistance with (red) and without (black) radiation.
			(c)~Comparison of $\delta R$ direct measurements (black) and double modulation technique (blue).
		}\label{DM}
	\end{figure*}

	In our measurements, see Fig.~\ref{DM}\,(a), the chopper frequency $f_\text{chopper} =23\,$Hz and a small bias current $I_\text{dc} = 1 - 50\,\mu$A is applied to the sample at much higher frequency $f_\text{current} = 433\,$Hz.
	The photoresistance is collected as a part of the signal that is modulated at both frequencies: 
	the first lock-in collects the total signal at higher frequency $f_\text{current}$, while its 	output is fed to the input of the second lock-in tuned to the lower frequency $f_\text{chopper}$. 
	The output signal of the second lock-in gives the value of $\delta R$. 
	
	The comparison of the direct measurements of $\delta R = R^\text{on} - R^\text{off}$ and using of the double modulation technique is shown in Fig.~\ref{DM}\,(b) and (c).
	It is seen that both methods coincide rather well. 
	
	\section{Transmittance, longitudinal resistance, and MIRO before and after room light illumination of the sample}
	\label{sec_light}

	\begin{figure}[h]
		\includegraphics[width=0.95\columnwidth]{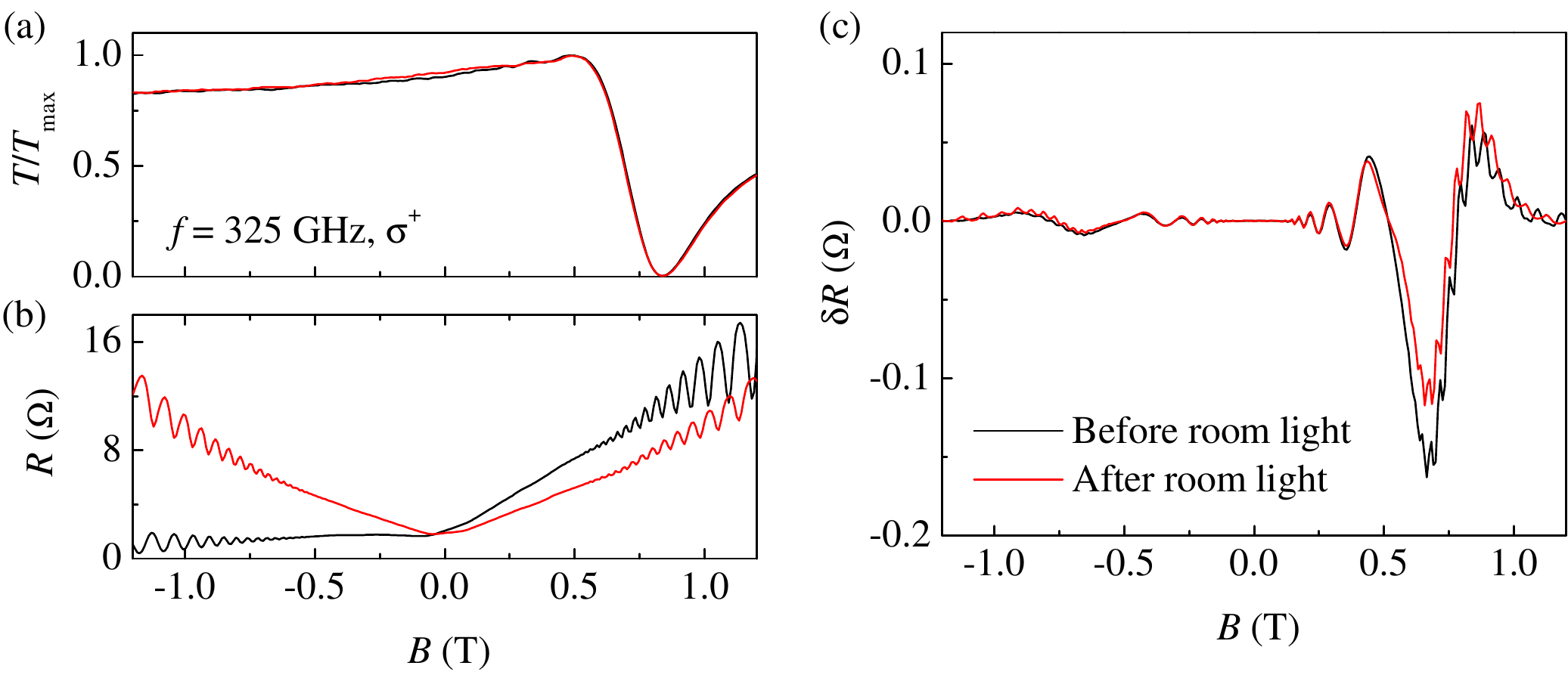}
		\caption{
			Magnetic field dependences of normalized transmittance $T/T_\text{max}(B)$~(a), longitudinal resistance $R(B)$~(b), and photoresistance $\delta R(B)$~(c) measured simultaneously at frequency $f = 325\,$GHz and for the right-hand circular polarization.
			Black and red colors correspond to the sample state before and after room light illumination of the sample, respectively.
			It is seen that in studied sample the room light illumination does not affect much the electron density (which would otherwise modify the shape of transmittance), the mobility, as well as the MIRO strength and high $B$-asymmetry reflecting high sensitivity of MIRO to radiation helicity.
		} \label{Before_after_light}
	\end{figure}

	\newpage
	\section{Technological design of the studied 2DES}
	\label{cross-section}

	\begin{figure}[h]
		\includegraphics[width=0.5\columnwidth]{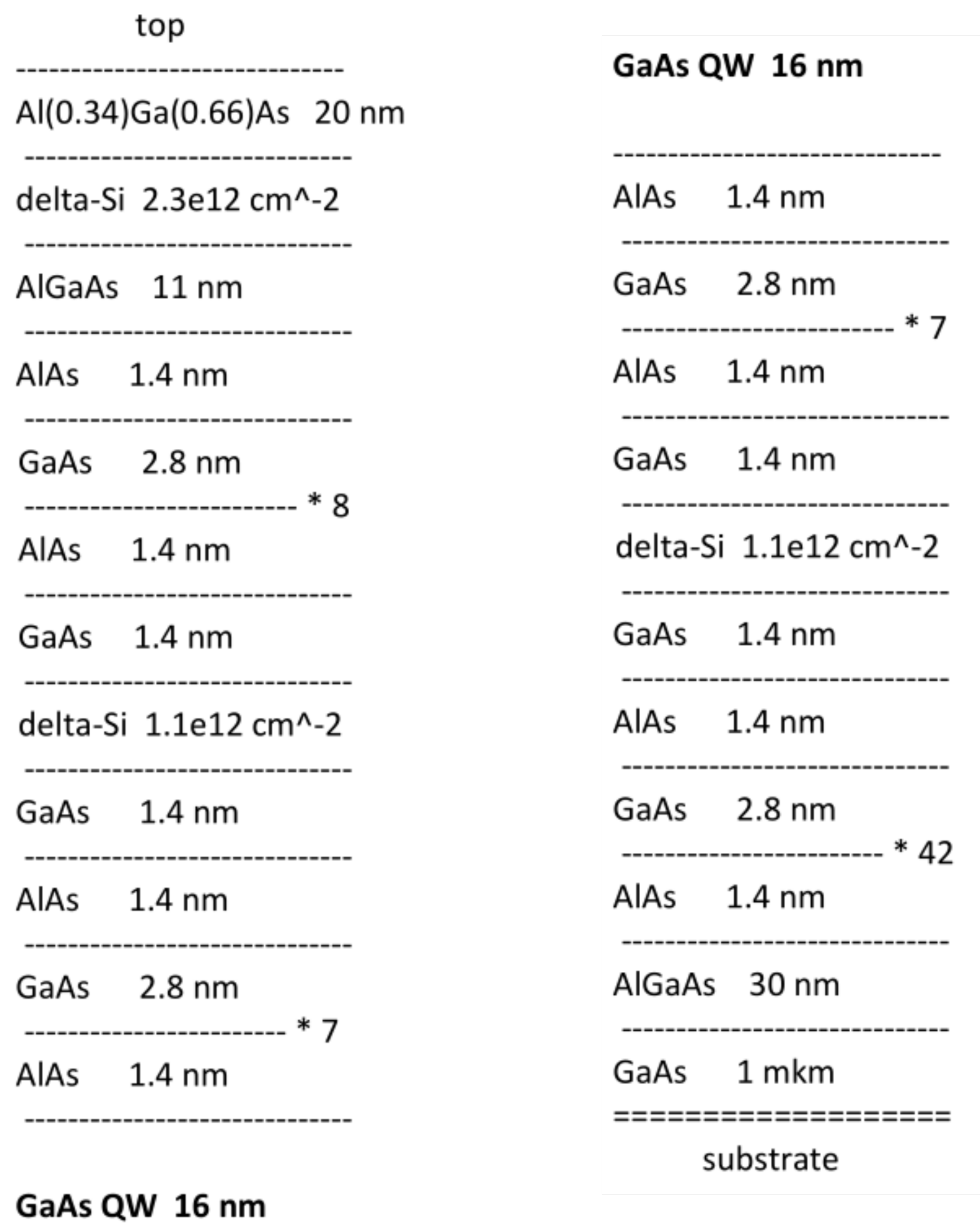}
		\caption{
			Heterostructure cross-section above quantum well with 2DES (left) and bellow it (right). 
		} \label{GaAs_16nm}
	\end{figure}

\end{document}